\documentclass[journal]{IEEEtran}
\usepackage{ifpdf}

\usepackage{amsmath}
\usepackage{algorithm}
\usepackage[noend]{algpseudocode}

\makeatletter
\def\BState{\State\hskip-\ALG@thistlm}
\makeatother

\usepackage{cite}

\ifCLASSINFOpdf
  \usepackage[pdftex]{graphicx}
   \DeclareGraphicsExtensions{.pdf,.jpeg,.png}
\else
\fi
\usepackage{lipsum}

\usepackage{amsmath}
\usepackage{amssymb}
\usepackage{flexisym}
\usepackage{array}
\newcolumntype{C}[1]{>{\centering\let\newline\\\arraybackslash\hspace{0pt}}m{#1}}

\usepackage{color}
\definecolor{dblue}{rgb}{0,0,0.8}

\usepackage{xcolor}

\usepackage{multirow}

\usepackage{subfig}
\usepackage{graphicx}
\usepackage{fixltx2e}

\usepackage{stfloats}
\usepackage{nomencl}

\usepackage{setspace}
\makenomenclature

\begin{document}
%


\title{Distributionally Robust Chance-Constrained Energy Management of Multi-Building Residential Apartment Complexes Using Wasserstein Metric}



%

\author{Hamed Haggi,~\IEEEmembership{Member,~IEEE}, James M. Fenton 
\thanks{The authors are with the University of Central Florida's Florida Solar Energy Center. Emails: hamed@ucf.edu, jfenton@fsec.ucf.edu.}
}

\maketitle

\begin{abstract}
The decreasing costs of photovoltaic (PV) systems and battery storage, alongside the rapid rise of electric vehicles (EVs), present a unique opportunity to revolutionize energy use in apartment complexes. Generating electricity via PV and batteries is currently cheaper and greener than relying on grid power, which is often expensive. Yet, residents in multi-building apartment complexes typically lack access to fast EV charging infrastructure. To this end, this paper investigates the feasibility and energy management of deploying commercial PV-powered battery storage and EV fast chargers within apartment complexes in Orlando, Florida, operated by complex owners. By modeling the complex as a grid-connected microgrid, it aims to meet residents' energy needs, provide backup power during emergencies, and introduce a profitable business model for property owners. To address PV power generation uncertainty, a distributionally robust chance-constrained optimization method using the Wasserstein metric is employed, ensuring robust and reliable operation. The techno-economic analysis reveals that EVs powered by PV and batteries are more cost-effective and environmentally friendly than gasoline vehicles that EV owners can save up to \$100 per month by saving on fuel costs. The results also show that integrating PV and battery systems reduces operational costs, lowers emissions, increases resilience, and supports EV adoption while offering a profitable business model for property owners. These findings highlight a practical and sustainable framework for advancing clean energy use in residential complexes.
\end{abstract}

\vspace{2mm}
\begin{IEEEkeywords}
Apartment Complex, Chance Constraints, Distributionally Robust Optimization, Energy Management, Smart Grids, Uncertainty, Wasserstein metric. 
\end{IEEEkeywords}

\IEEEpeerreviewmaketitle

\section{Introduction}
\IEEEPARstart{T}{he} global energy sector is undergoing significant changes due to the rapid growth of distributed energy resources (DERs), attributed to their technological advancements, economic viability, and environmental benefits. As DERs such as solar photovoltaic (PV) systems, battery energy storage (BES) technologies, and electric vehicles (EVs) continue to proliferate, they are reshaping traditional power and energy systems by promoting decentralized energy generation, enhancing grid resilience, and empowering consumers with greater control over their energy use. Thus, the integration of these DERs can significantly contribute to power system decarbonization and transportation electrification.

A major opportunity for DERs lies in the building sector, which is responsible for nearly 73\% of total electricity consumption in the United States, with 38\% attributed to residential buildings and 35\% to commercial buildings, according to the U.S. Energy Information Administration (EIA) \cite{EIAbuildings}. This highlights the critical role that buildings play in the energy landscape and the potential impact of integrating DERs into these areas. By adopting technologies such as PV systems, BES technologies, and EV charging infrastructure, buildings can transition toward becoming net-zero energy, where on-site renewable energy generation meets or exceeds their energy demands. Effectively operating these DERs in a cost-efficient manner while addressing techno-economic challenges is crucial for the building sector to minimize operational costs and energy losses while maximizing the use of affordable and renewable green energy.

Building energy management topics considering smart technologies were widely discussed in the literature. For instance, authors of reference \cite{mariano2021review} presented a comprehensive review of strategies for energy management of residential and non-residential buildings from building topology, optimization techniques, building subsystems, management strategies, energy efficiency, etc. perspectives. Han et al. \cite{han2023home} presented a review on home energy management systems (HEMS), small appliances, concepts, and scheduling purposes. Focusing on the technical side, Elkadeem et al. \cite{elkadeem2023optimal} proposed a planning framework considering PV, combined heat and power, battery, and EVs for grid-connected multi-residential buildings with the goal of minimizing the overall cost.  Authors of \cite{comodi2015multi} only presented operational results for a small building with six units (like townhouse buildings), PV, battery, and thermal storage to satisfy hot water summer demand. Foroozandeh et al. in \cite{foroozandeh2022single} proposed an intelligent energy management framework, optimizing contract power decisions in smart buildings to reduce costs effectively using PV, battery, and EVs. Tamashiro et al. \cite{tamashiro2023optimal} proposed a smart apartment building optimization model using PV, battery, thermal storage, and fuel cell units. The model predictive control (MPC) algorithm was used to minimize two objectives; the cost and carbon emission. In another work, \cite{nagarsheth2024energy} studied a fuzzy logic-based method for the energy management of residential apartment buildings in Quebec, Canada. The units are centrally optimized to minimize the cost and provide thermal comfort. \cite{rezaei2020optimal} optimized a multi-unit residential building considering localized PV, energy storage, and EVs. This centralized energy management system controls the comfort parameters while responding to price variations. Authors of \cite{saberi2021data} proposed a data-driven distributional robust hierarchical coordination for building's energy management with a focus on heating, ventilation, air conditioning (HVAC) systems, and small home appliances. In another uncertainty-related work, \cite{yamamoto2024mpc} studied a two-stage MPC-based robust optimization of smart apartment buildings considering the uncertainty of PV generation and electricity demand which is based on an interval partitioned robust method. 

Urban residential complexes, particularly multi-building residential apartment complexes, present a unique opportunity to capitalize on these developments. However, previous research works did not focus on multi-building residential apartment complexes that include both residential and commercial loads (including commercial buildings load, EV demand, and commercial PV + battery storage), treating all loads the same. They often used a single pricing structure, like fixed or time-of-use (TOU) rates, for scheduling, without considering the different needs of residential loads, commercial buildings, and EV charging stations, and regulations. Also, many studies ignored the uncertainty of renewable energy, which can greatly affect the financial benefits of these systems. Previous research also mainly considered normal operation, while this work includes resilience in the design and operation, ensuring the system's performance against unexpected issues. Finally, earlier studies failed to consider real-world data from an existing complex to improve accuracy and practical application, and their techno-economic feasibility. To this end, the main contributions of this paper are:

\begin{itemize}
    \item Proposed a resilience-oriented model for energy management of grid-connected multi-building residential apartment complexes equipped with commercial PV, commercial BES, and EV charging stations. The model is decomposed for residential and commercial building types with appropriate demand and pricing structures. 
    \item Developed and implemented a distributionally robust chance-constrained method using the Wasserstein metric to address the PV power uncertainty and through sensitivity analysis the impact of risk is considered on operational cost. Furthermore, empirical input data such as PV power data, electric load, and two price structures were utilized for an existing apartment complex in the city of Orlando. 
\end{itemize}

The rest of the paper is organized as follows: Section \ref{problem_formulation_APT} will provide information regarding the problem formulation. Section \ref{Results_analysis} will provide simulation inputs and appropriate parameters. Finally, Section \ref{conclusionn} concludes the paper.




\section{Problem Formulation}\label{problem_formulation_APT}
This section provides detailed problem formulation of grid-connected multi-building residential apartment complexes' energy management with PV, BES, and EV fast chargers within the community in both deterministic and robust ways considering renewable energy uncertainty. 
\subsection{Deterministic Formulation of the Problem}
The proposed framework's objective function, Equation (\ref{OF}), is to minimize the total cost from the apartment complex owner's perspective. To simplify the problem formulation, we have not separated it into normal and emergency operation modes. Prior to major disruptions occurring, all the equations remain valid as presented, except for the load-shedding terms, which should be ignored or set to zero.
\begin{equation}\label{OF}
\begin{split}
     \min \sum_{t=1}^{T} [ & \; C_{t}^{Com} + C_{t}^{Res} + C_{t}^{PV} + C_{t}^{Res_{Shd}}\\& + C_{t}^{Com_{Shd}}  + C_{t}^{ResCom_{Shd}} ]
\end{split}
\end{equation}
\begin{equation}\label{Cost_commercial}
    C_{t}^{Com} = [P_{t}^{GB_{C}} - P_{t}^{GS_{C}}] \;.\; \lambda_{t}^{C} 
\end{equation}
\begin{equation}\label{Cost_residential}
    C_{t}^{Res} = P_{t}^{GB_{R}} \;.\; \lambda_{t}^{R} 
\end{equation}
\begin{equation}\label{Cost_PV}
    C_{t}^{PV} = P_{t}^{PV}  \;.\; \lambda_{t}^{PV} 
\end{equation}
\begin{equation}\label{LSCost_residential}
    C_{t}^{Res_{Shd}} = P_{t}^{R,Shd}  \;.\; voll^{R,Shd}
\end{equation}
\begin{equation}\label{LSCost_commercial}
    C_{t}^{Com_{Shd}} = P_{t}^{C,Shd}  \;.\; voll^{C,Shd}
\end{equation}
\begin{equation}\label{LSCost_rescom}
    C_{t}^{ResCom_{Shd}} = P_{t}^{RC,Shd}  \;.\; voll^{RC,Shd}
\end{equation}
Let us define $P^{GB_{C}}_t$, $P^{GS_{C}}_t$, $P^{GB_{R}}_t$ as buying and selling power from/to the grid for commercial purposes, and purchasing power for residential load, respectively. $P^{PV}_t$, $P^{R_{Shd}}_t$, $P^{C_{Shd}}_t$, and $P^{RC_{Shd}}_t$, are PV power, load shedding variables for residential, commercial, and common electricity in residential buildings.  Equation (\ref{OF}) consists of 6 terms with formulations presented in Equations (\ref{Cost_commercial})-(\ref{LSCost_rescom}) where the terms refer to the exchanged power cost of commercial buildings with grid, purchasing power by residents, cost of PV, and load shedding penalty terms for residential load, common residential buildings load, and commercial buildings load, respectively.
\begin{equation}\label{grid_buy_limit_residential}
    0 \le P_{t}^{GB_{R}} \le P^{G,max}
\end{equation}
\begin{equation}\label{grid_buy_limit_commercial}
    0 \le P_{t}^{GB_{C}} , P_{t}^{GS_{C}} \le P^{G,max}
\end{equation}
Equation (\ref{grid_buy_limit_residential}) shows the power purchase limit (defined as $P^{G, {max}}$) for residents while Equation (\ref{grid_buy_limit_commercial}) shows the buying/selling power of apartment complex for its commercial buildings from/to the grid.
\begin{equation}\label{PV_limit}
    0 \le P_{t}^{PV_{Res}}, P_{t}^{PV_{Com}}, P_{t}^{PV}  \le P_{t}^{PV,max}
\end{equation}
\begin{equation}\label{PV_sum}
    P_{t}^{PV_{Res}} + P_{t}^{PV_{Com}} = P_{t}^{PV}
\end{equation}
\begin{equation}\label{PV_crtl}
     P_{t}^{PV} + P_{t}^{PV_{Crtl}} - P_{t}^{PV,max} = 0
\end{equation}
Equations (\ref{PV_limit}) to (\ref{PV_crtl}) show the PV power limit ($P_t^{PV,max}$), and curtailed PV power ($P_t^{PV_{Crtl}}$) constraint, respectively. 
\begin{equation}\label{BES_charge}
    0 \le P_{t}^{BS,ch}\;.\; \eta_{BS} \le P^{BS,Cap}
\end{equation}
\begin{equation}\label{BES_discharge}
    0 \le P_{t}^{BS,dis}\;/\; \eta_{BS} \le P^{BS,Cap}
\end{equation}
\begin{equation}\label{BES_SOC}
    SOC_{t}^{BS} = SOC_{t-1}^{BS} + (P_{t}^{BS,ch}\;.\; \eta_{BS} - P_{t}^{BS,dis}\;/\; \eta_{BS}).\Delta t
\end{equation}
\begin{equation}\label{BES_SOC_limit}
    SOC^{BS,min} \le SOC_{t}^{BS} \le SOC^{BS,max}
\end{equation}
Considering $P_{t}^{BS,ch}$, $P_{t}^{BS,dis}$, $SOC_{t}^{BS}$, $P^{BS,Cap}$, and $SOC^{BS, max/min}$ as charging and discharging power of BES, maximum power rating and limits on state of charge of battery, set of Equations (\ref{BES_charge})-(\ref{BES_SOC_limit}) refer to battery energy storage constraints (to save space we only present one set of equations. These equations must be considered for both residential and commercial purposes) where the first two equations refer to charging and discharging power limits of the battery respectively. Equation (\ref{BES_SOC}) shows the energy state of charge of the battery and Equation (\ref{BES_SOC_limit}) put boundaries on this variable. To decompose the formulation we have: 

\begin{equation}\label{BES_SOC_limit}
    P_{t}^{BS,ch_{R}} + P_{t}^{BS,ch_{C}} = P_{t}^{BS,ch}
\end{equation}
where indexes $R$ and $C$ in variables refer to residential and commercial. 
\begin{equation}\label{EV_charge}
    0 \le P_{t,i}^{EV,ch}\;.\; \eta_{EV} \le P_{i}^{EV Cap}
\end{equation}
\begin{equation}\label{EV_discharge}
    0 \le P_{t,i}^{EV,dis}\;/\; \eta_{EV} \le P_{i}^{EV Cap}
\end{equation}
\begin{equation}\label{EV_SOC}
    SOC_{t,i}^{EV} = SOC_{t-1,i}^{EV} + (P_{t,i}^{EV,ch}\;.\; \eta_{EV} - P_{t,i}^{EV,dis}\;/\; \eta_{EV}).\Delta t
\end{equation}
\begin{equation}\label{EV_SOC_limit}
    SOC_{i}^{EV,min} \le SOC_{t,i}^{EV} \le SOC_{i}^{EV,max}
\end{equation}
\begin{equation}\label{EV_SOC_arrival}
     SOC_{t = Tarrival,i}^{EV} = SOC_{i}^{EV, arrival} 
\end{equation}
\begin{equation}\label{EV_SOC_departure}
     SOC_{t = Tdeparture,i}^{EV} = SOC_{i}^{EV, departure} 
\end{equation}

Equations (\ref{EV_charge})-(\ref{EV_SOC_departure}) show EVs constraints whereas Equations (\ref{EV_charge}) and (\ref{BES_discharge}) show the charging and discharging of EVs at parking stations (defined as $P_{t,i}^{EV,ch}$ and $P_{t,i}^{EV,dis}$). Equation (\ref{EV_SOC}) shows the state of charge of the cars' batteries (defined as $SOC_{t,i}^{EV}$) and Equation (\ref{EV_SOC_limit}) shows the minimum and maximum state of charge of these batteries. Finally, each EV arrives at a parking spot equipped with a fast charger with initial SOC ($SOC_{i}^{EV, arrival} $) which is modeled by Equation (\ref{EV_SOC_arrival}). The departure SOC of EVs leaving the charging station ($SOC_{i}^{EV, departure}$) is also modeled by Equation (\ref{EV_SOC_departure}).

\begin{equation}\label{Pbalance_residential}
    P_{t}^{GB_{R}} - P_{t}^{load_{R}} + P_{t}^{PV_{Res}} - P_{t}^{BS,ch_{R}} +  P_{t}^{BS,dis_{R}} + P_{t}^{R,Shd} = 0
\end{equation}
\begin{equation}\label{Pbalance_commercial}
\begin{split}
    & P_{t}^{GB_{C}} - P_{t}^{GS_{C}} - P_{t}^{load_{C}} - P_{t}^{load_{RC}} + P_{t}^{PV_{Com}} - P_{t}^{BS,ch_{C}} \\& +  P_{t}^{BS,dis_{C}} + P_{t}^{C,Shd} + P_{t}^{RC,Shd} - \sum_{i=1}^{N_{EV}}(P_{t}^{EV,ch}) = 0
\end{split}
\end{equation}

Considering $P_{t}^{load_{R}}$ and $P_{t}^{load_{C}}$ as residential and commercial buildings' electricity consumption, both residential and commercial power balance constraints are presented in Equations (\ref{Pbalance_residential}) and (\ref{Pbalance_commercial}), respectively.

\subsection{Distributionally Robust Chance Constrained Optimization Using Wasserstein Metric (DRCC-WM)}\label{DRCC_formula}

In this section, a distributionally robust chance-constrained optimization framework using the Wasserstein metric (DRCC-WM) \cite{li2022wasserstein} is introduced. The uncertainty in PV generation is modeled using an ambiguity set based on the Wasserstein metric. The empirical distribution $\hat{P}_N$ of the uncertain PV generation $\xi$ is constructed from $N$ observed samples:

\begin{equation}
\hat{P}_N = \frac{1}{N} \sum_{k=1}^N \delta_{\xi^{(k)}}
\end{equation}

where $\delta_{\xi^{(k)}}$ is the Dirac delta function centered at the $k$-th observed sample $\xi^{(k)}$. This empirical distribution assigns equal probability to each observed sample, effectively capturing the observed variability in PV generation.

Aiming to ensure that the operational constraints are satisfied with high probability for all plausible distributions of the uncertain PV generation within a specified ambiguity set $\mathcal{P}$:

\begin{equation}
\inf_{P \in \mathcal{P}} P[g(x, \xi) \leq 0] \geq 1 - \alpha
\end{equation}

where $g(x, \xi)$ represents the constraint function dependent on decision variables $x$ and uncertainty $\xi$, and $\alpha \in (0,1)$ is the acceptable risk level. The infimum over $P \in \mathcal{P}$ ensures robustness against the worst-case distribution within the ambiguity set. The Wasserstein metric $W(Q_1, Q_2)$ quantifies the distance between two probability distributions $Q_1$ and $Q_2$ over the uncertainty space $\mathcal{E}$:

\begin{equation}
W(Q_1, Q_2) = \inf_{\Pi} \left\{ \int_{\mathcal{E}^2} \|\xi_1 - \xi_2\| \; \Pi(d\xi_1, d\xi_2) \right\}
\end{equation}

where $\Pi$ denotes the set of all joint distributions on $\mathcal{E} \times \mathcal{E}$ with marginals $Q_1$ and $Q_2$. The Wasserstein metric measures the minimal expected transportation cost of moving mass from $Q_1$ to $Q_2$, using the Euclidean distance $\|\xi_1 - \xi_2\|$ as the cost function. Let us define the ambiguity set $\mathcal{P}$ as a Wasserstein ball around the empirical distribution $\hat{P}_N$ with radius $\epsilon(N)$:

\begin{equation}
\hat{P}_N  = \left\{ P \in \mathcal{P}(\mathcal{E}) : W(P, \hat{P}_N) \leq \epsilon(N) \right\}
\end{equation}

where $\mathcal{P}(\mathcal{E})$ denotes the set of all probability distributions on $\mathcal{E}$. This includes all distributions that are within a Wasserstein distance $\epsilon(N)$ from the empirical distribution, accounting for statistical estimation errors due to a finite sample size. $\epsilon(N)$, the radius, in terms of the constant $C$ and confidence level $1 - \beta$, where $C$ is the diameter of uncertainty and a positive constant depending on the distributions $P$ and $Q$.


\begin{equation}
\epsilon(N) = C\sqrt{\frac{1}{N} \log \left(\frac{1}{1-\beta}\right)}
\end{equation}

This expression aligns with Equation and ensures that the ambiguity set $\mathcal{P}$ contains the true distribution with probability at least $1 - \beta$.
Since the true distribution $P$ is unknown, $C$ can be estimated using the empirical data:

\begin{equation}
\begin{split}
C &\leq 2 \inf_{\alpha > 0} \left( \frac{1}{2\alpha} \left( 1 + \ln \mathbb{E}_P\left[e^{\alpha \|\xi - \hat{\mu}\|^2}\right] \right) \right)^{1/2} \\
&\approx 2 \inf_{\alpha > 0} \left( \frac{1}{2\alpha} \left( 1 + \ln \left( \frac{1}{N} \sum_{k=1}^N e^{\alpha \|\xi^{(k)} - \hat{\mu}\|^2} \right) \right) \right)^{1/2},
\end{split}
\end{equation}

where $\hat{\mu} = \frac{1}{N} \sum_{k=1}^N \xi^{(k)}$ is the empirical mean of the samples. The approximation replaces the expectation $\mathbb{E}_P$ with the empirical average over the observed data, allowing for practical computation of $C$.

\subsection{Tractable Compact Reformulation of DRCC-WM}
The compact form of optimization problem considering uncertain parameter is presented in Equations \ref{compact_OF} to \ref{compact_const}.
\begin{equation}\label{compact_OF}
\min_{x} \left\{ C_0(x) + \sup_{P \in \mathcal{P}} \mathbb{E}_{P} \left[ C_1(x, \xi) \right] \right\}
\end{equation}
\begin{equation}\label{compact_DRCC}
\inf_{P \in \mathcal{P}} \, P\left[ H_k(x, \xi) \leq 0 \right] \geq 1 - \alpha, \quad \text{for } k = 1, \dots, p
\end{equation}
\begin{equation}\label{compact_uncertainOF}
C_1(x, \xi) = b(x)^\top \xi
\end{equation}
\begin{equation}\label{compact_const}
H_k(x, \xi) = h_{0k}(x) + a_k(x)^\top \xi
\end{equation}

The objective function minimizes the sum of a deterministic cost component $C_0(x)$ and the worst-case expected value of an uncertain cost  $C_1(x, \xi)$ over all probability distributions $P$ within an ambiguity set \( \mathcal{P} \) defined by the Wasserstein ball. The uncertain constraints are addressed by ensuring they hold high probability for all distributions in \( \mathcal{P} \). 

\begin{equation}\label{compact_Re_OF}
\min_{x} \left\{ C_0(x) + \frac{1}{N} \sum_{i=1}^{N} b(x)^\top \xi_i + \delta \, \| b(x) \|_* \right\}
\end{equation}
\begin{equation}\label{compact_Re_equivalentCC}
h_{0k}(x) + a_k(x)^\top \xi_i + \delta \, \| a_k(x) \|_* \leq 0, \quad
\begin{cases}
i = 1, \dots, N \\
k = 1, \dots, p
\end{cases}
\end{equation}

By assuming affine dependence on the uncertain parameters \( \xi \), the infinite-dimensional optimization over probability distributions is transformed into a tractable problem. The worst-case expectations are replaced with empirical averages computed from observed data samples, augmented by terms involving the dual norms of the coefficient vectors and the Wasserstein radius \( \delta \). This compact formulation simplifies the original problem while maintaining robustness against distributional uncertainty, making it suitable for efficient computational implementation in practical applications.

\begin{figure}
\centering
\footnotesize
\captionsetup{singlelinecheck=false,font={footnotesize}}
	\includegraphics[width=3.3in,height=1.6in]{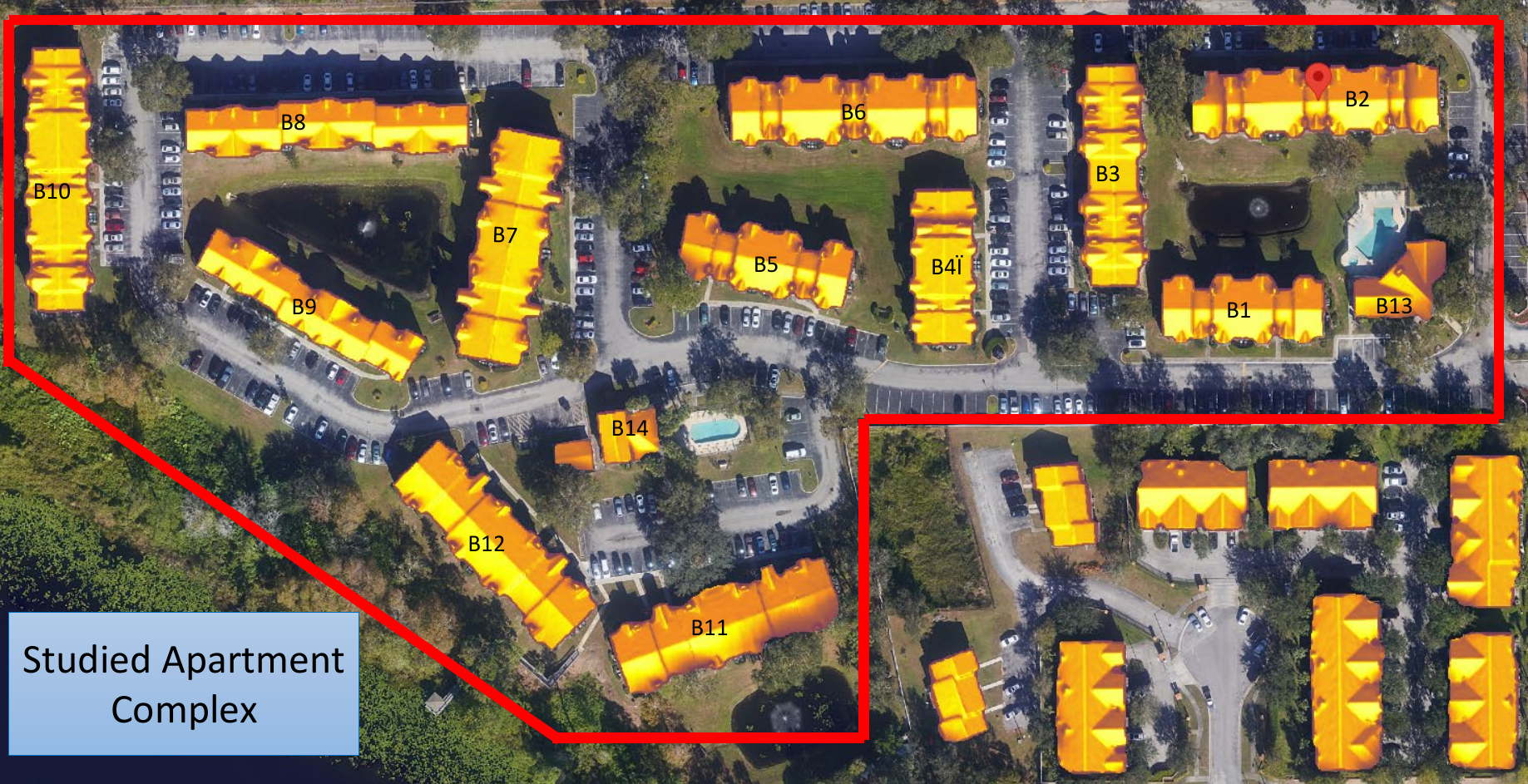}
	\caption{Potential useful solar panel installation areas for studied apartment complex (Google Sunroof).}
    \label{Googlesunroof}
\end{figure}

\section{Simulation Results and Analysis}\label{Results_analysis}
\begin{figure*}
\center
\footnotesize
\captionsetup{font={footnotesize}}
 \begin{tabular}{cc}
 \subfloat[Time of use electricity rates.] {\includegraphics[height=1.6in,width=2.1in]{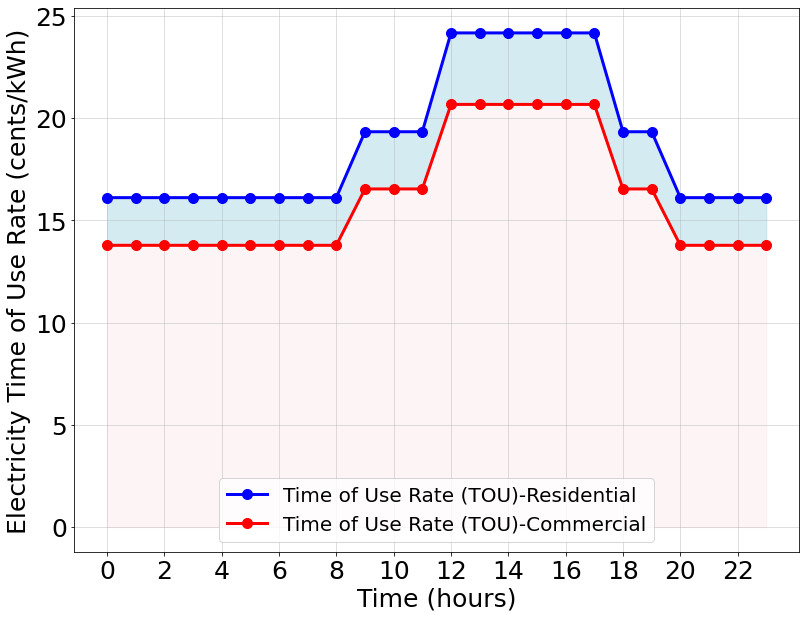}\label{TOU_rates} }
 \subfloat[Apartment complex demand.]{\includegraphics[height=1.6in,width=2.1in]{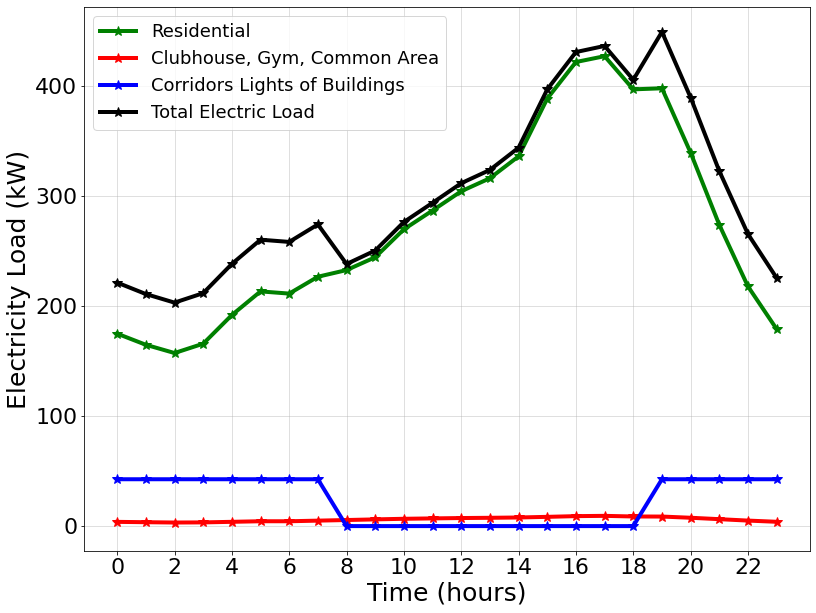}\label{Total_demand} }
  \subfloat[Aggregated rooftop solar power.] {\includegraphics[height=1.6in,width=2.1in]{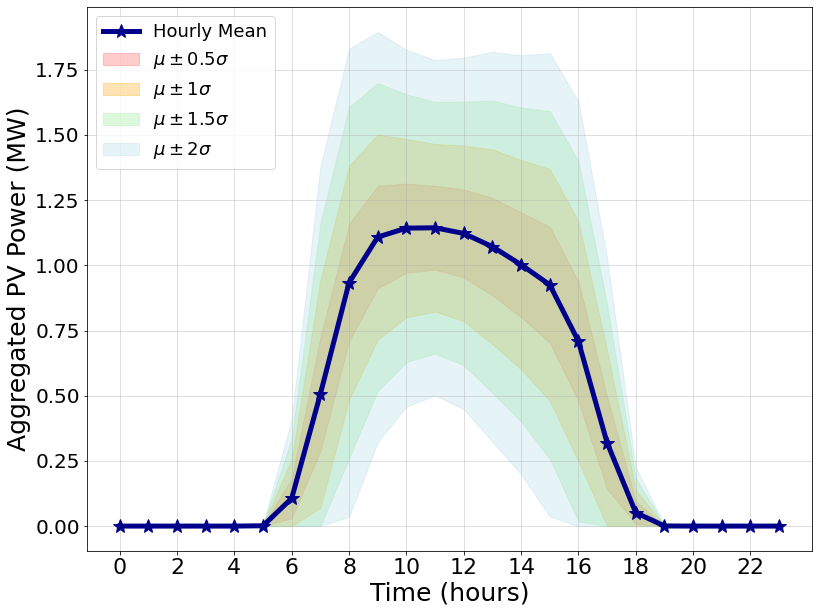}\label{PV_DRO} }
 \end{tabular}
 \caption{Electricity TOU rates, apartment complex's electricity demand, and aggregated PV power for 14 complex buildings with different confidence levels.}
\label{Fig. Voltage Convergence CaseI}
\end{figure*}

The proposed DRCC-WM apartment complex energy management problem is validated by testing on a real apartment complex located in the city of Orlando, Florida. As it can be seen from Figure \ref{Googlesunroof}, the apartment complex has 14 buildings (12 residential and 2 commercial buildings). Residential buildings are one-, two-, and three-bedroom units. More information regarding building number, number of units, and commercial building types is presented in Table \ref{Building_parameters}. The TOU electricity rates for residential and commercial buildings are presented in Figure \ref{TOU_rates}. These rates include energy charge rate, fuel charge rate, asset utilization rate, and fixed cost associated with both residential and commercial (obtained from actual bills). Figure \ref{Total_demand} shows the total residents' demand, clubhouse gym and common area electricity consumption, and corridor lights of the apartment complex buildings. Using the Google Project Sunroof tool \cite{GoogleSunroofProject}, optimal areas are identified to install rooftop solar panels based on solar irradiance and shaded areas which are shown in Figure \ref{Googlesunroof}. The aggregated PV power is 1.54 MW for all 14 buildings presented in Figure \ref{PV_DRO} and its appropriate commercial levelized cost of electricity is obtained from \cite{NRELATB}. To make the apartment complex grid independent, a 4-hour Li-ion battery with an 800 kW power rating and an efficiency of 90\% is installed. In this complex, five fast chargers are considered for residents' EVs and the arrival and departure time and SOC are presented in Table \ref{EV_parameters}. EVs are Tesla Model 3 Long Range \cite{TeslaModel3_EVDatabase} with 330 miles range, 11 kW power with a battery capacity of approximately 75 kWh, 90\% battery efficiency, and level 2 chargers. 

\begin{table}[]
\footnotesize 
\centering
\captionsetup{labelsep=space,font={footnotesize,sc}}
\caption{ \\ Key Apartment Complex Buildings Information.}
\label{Building_parameters}
\begin{tabular}{|c|c|c|c|}
\hline 
Type                        & Bedrooms                                                     & Buildings        & Units          \\ \hline \hline
\multirow{3}{*}{Residential} &
  One-Bed &
  \begin{tabular}[c]{@{}c@{}}B1, B2, B4, B5\\  B6, B10\end{tabular} &
  \begin{tabular}[c]{@{}c@{}}16, 36, 24,  16\\ 36, 36\end{tabular} \\ \cline{2-4} 
                            & \begin{tabular}[c]{@{}c@{}}\ Two-Bed \\ \end{tabular}          & B3, B7, B11, B12 & 24, 24, 16, 16 \\ \cline{2-4} 
                            & \begin{tabular}[c]{@{}c@{}}Three-Bed\\ \end{tabular}        & B8, B9           & 20, 36         \\ \hline
\multirow{2}{*}{Commercial} & \begin{tabular}[c]{@{}c@{}}Clubhouse\\ Mailroom\end{tabular} & B13              & 1              \\ \cline{2-4} 
                            & \begin{tabular}[c]{@{}c@{}}Gym\\ Pools\end{tabular}          & B14              & 1              \\ \hline 
\end{tabular}
\end{table}

\begin{table}[]
\centering
\footnotesize 
\centering
\captionsetup{labelsep=space,font={footnotesize,sc}}
\caption{ \\ Electric Vehicles Technical Parameters.}
\label{EV_parameters}
\begin{tabular}{|c|c|c|c|c|c|}
\hline

 &
  \begin{tabular}[c]{@{}c@{}}Energy \\ Capacity\\ (kWh)\end{tabular}  & 
  \begin{tabular}[c]{@{}c@{}}Arrival \\ Time\\ (h)\end{tabular} &
  \begin{tabular}[c]{@{}c@{}}Departure \\ Time\\ (h)\end{tabular} &
  \begin{tabular}[c]{@{}c@{}}SOC \\ Arrival\\ (\%)\end{tabular} &
  \begin{tabular}[c]{@{}c@{}}SOC \\ Departure\\ (\%)\end{tabular} \\ \hline \hline
EV 1 & 75 & 5  & 12 & 20 & 95  \\ \hline
EV 2 & 75 & 9  & 15 & 20 & 100 \\ \hline
EV 3 & 75 & 15 & 21 & 20 & 90  \\ \hline
EV 4 & 75 & 6  & 11 & 40 & 80  \\ \hline
EV 5 & 75 & 18 & 23 & 30 & 90  \\ \hline
\end{tabular}
\end{table}

The simulations are carried out on a PC with an Intel(R) Xeon(R) W-1250 CPU of 3.31 GHz and 64 GB RAM. The proposed framework is coded in Python/Pyomo and solved using the Gurobi solver.
\subsection{Smart Energy Management Optimization Results}
\begin{figure*}
\center
\footnotesize
\captionsetup{font={footnotesize}}
 \begin{tabular}{cc}
 \subfloat[C vs parameter eta optimization result.] {\includegraphics[height=1.6in,width=2.1in]{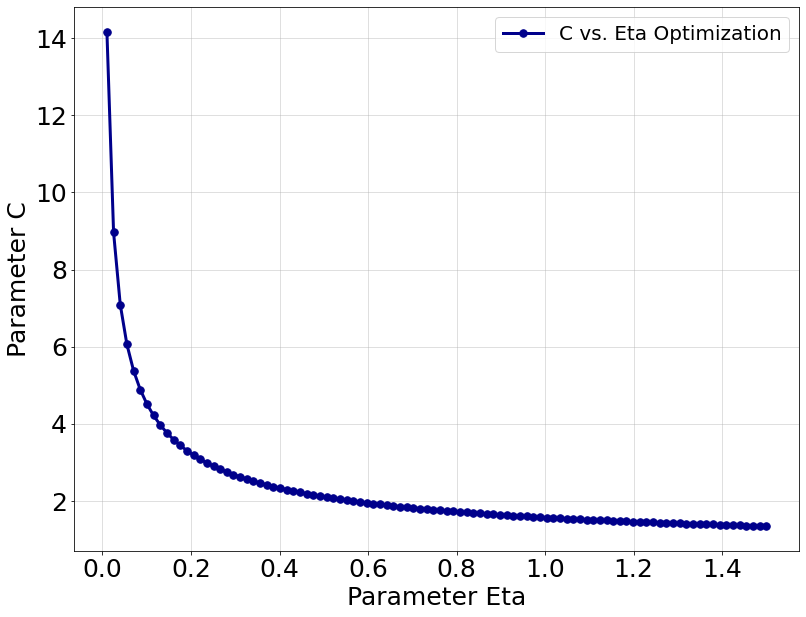}\label{CvsEta} }
 \subfloat[Energy scheduling of apartment complex.]{\includegraphics[height=1.6in,width=2.1in]{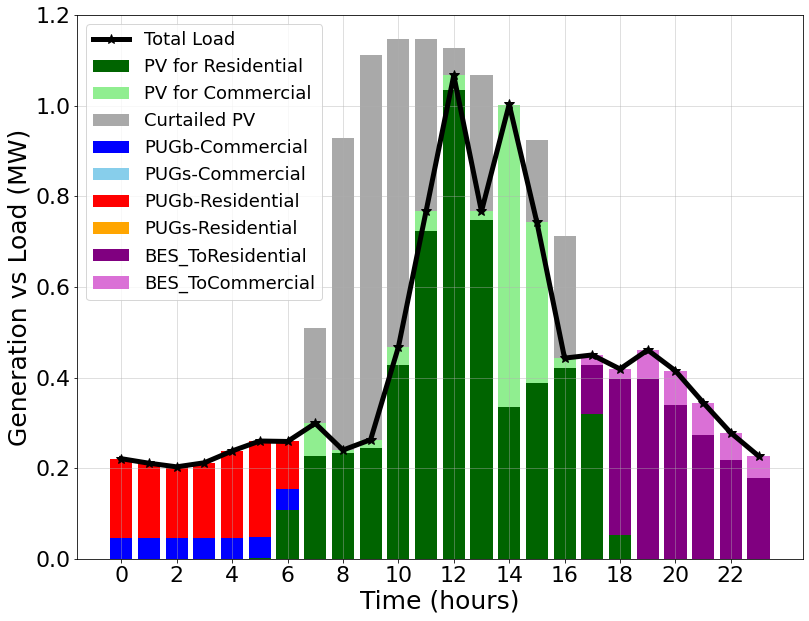}\label{Genloadd}} 
  \subfloat[Battery storage scheduling and SOC.] {\includegraphics[height=1.6in,width=2.2in]{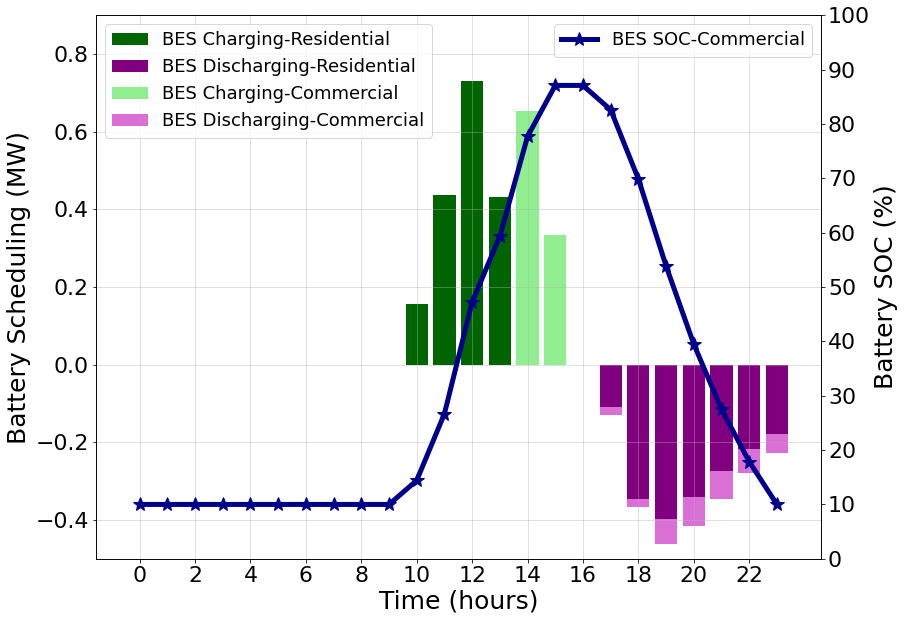}\label{Schedule_Battery}}
 \end{tabular}
 \caption{Smart energy management of apartment complex with PV, Battery, and EVs for the confidence level of 95\%.}
\label{schedule_PVBES}
\end{figure*}

\begin{figure}
\centering
\footnotesize
\captionsetup{singlelinecheck=false,font={footnotesize}}
	\includegraphics[width=3.4in,height=1.8in]{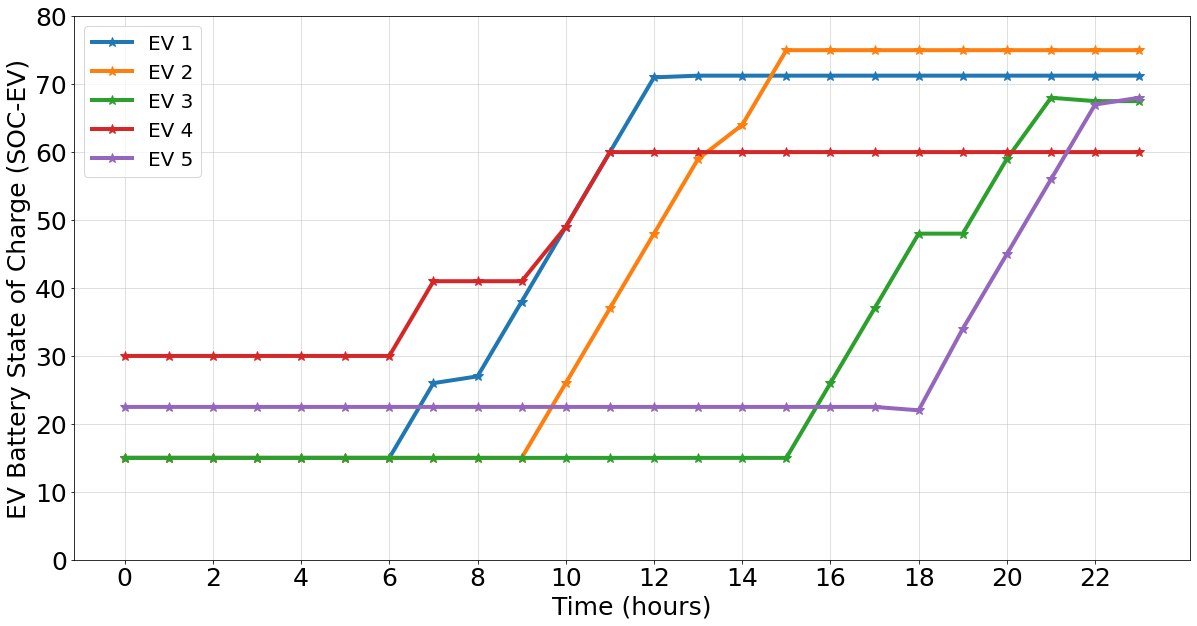}
	\caption{State of Charge of EVs' Batteries.}
    \label{EVResults}
\end{figure}

The results for smart energy management of multi-building residential apartment complex with PV, battery, and EVs for a confidence level of 95\% (risk level = 5\%) are presented in Figure \ref{schedule_PVBES} and Figure \ref{EVResults}. Figure \ref{CvsEta} shows the optimization result of the Wasserstein Constant $C$ with respect to the parameter $\eta$ which has an impact on the accuracy of approximations and predictions. Based on this result and computational aspects we chose $C = 1.36$ as there is not much change even if we extend the range of parameter $\eta$ to higher numbers.

Figure \ref{Genloadd} depicts the energy scheduling of the system. As observed, during the first five hours, the apartment complex purchases electricity from the grid to meet its demand. However, as PV power becomes available, the system prefers to utilize the less expensive PV-generated electricity over grid electricity. Since the PV power exceeds the electricity demand, the commercial battery begins charging to be utilized during nighttime, aiming to achieve grid independence and reduce costs. The battery charging profile and its utilization of PV power for both residential and commercial purposes are presented in Figure \ref{Genloadd}, and the state of charge (SOC) of the apartment complex's battery is shown in Figure \ref{Schedule_Battery}.

During nighttime, the energy stored in the battery can be used to offset the residents' electricity consumption and also provide electricity to EV owners through fast chargers. The EV batteries' SOCs are shown in Figure \ref{EVResults}. In the state of Florida, third-party entities are allowed to sell electricity to customers through fast chargers. Since this design focuses on this business model, it is worthwhile to compare the fuel costs between a gasoline vehicle and an EV that can be charged in these apartment complexes. Assuming the Tesla Model 3 has specs of 330 miles and a 75 kWh battery, it results in an efficiency of 4.4 miles per kWh. The average gasoline vehicle achieves 24.2 miles per gallon (mpg) \cite{USDOEEERE}. Therefore, the gasoline equivalent in dollars per gallon can be calculated by multiplying the price of electricity in dollars per kWh by the ratio of 24.2 mpg to 4.4 miles per kWh. The LCOE for commercial rooftop solar is 8.8 cents per kWh. This is equivalent to \$0.49 per gallon of gasoline, calculated as 0.088 (\$/kWh)  ×  24.2 (mile/gallon)  ÷  4.4 (mile/kWh) = 0.49 (mile/gallon). Considering that the average price for gasoline in 2024 is \$3.56 per gallon in 2024 \cite{GasFLPrice}, the gasoline equivalent cost of rooftop PV is approximately 14\% of the cost of gasoline which is way cheaper. Even if we calculate the maximum commercial TOU rate which is 21 cents per kWh, the gasoline equivalent will be \$1.6/gallon which is 33\% of current gasoline prices (67\% cheaper).

\subsection{Operational Cost Sensitivity Analysis for Various Parameter $\alpha$ Values for Deterministic, CC, and DRCC-WM}
\begin{figure}
\centering
\footnotesize
\captionsetup{singlelinecheck=false,font={footnotesize}}
	\includegraphics[width=3.4in,height=1.85in]{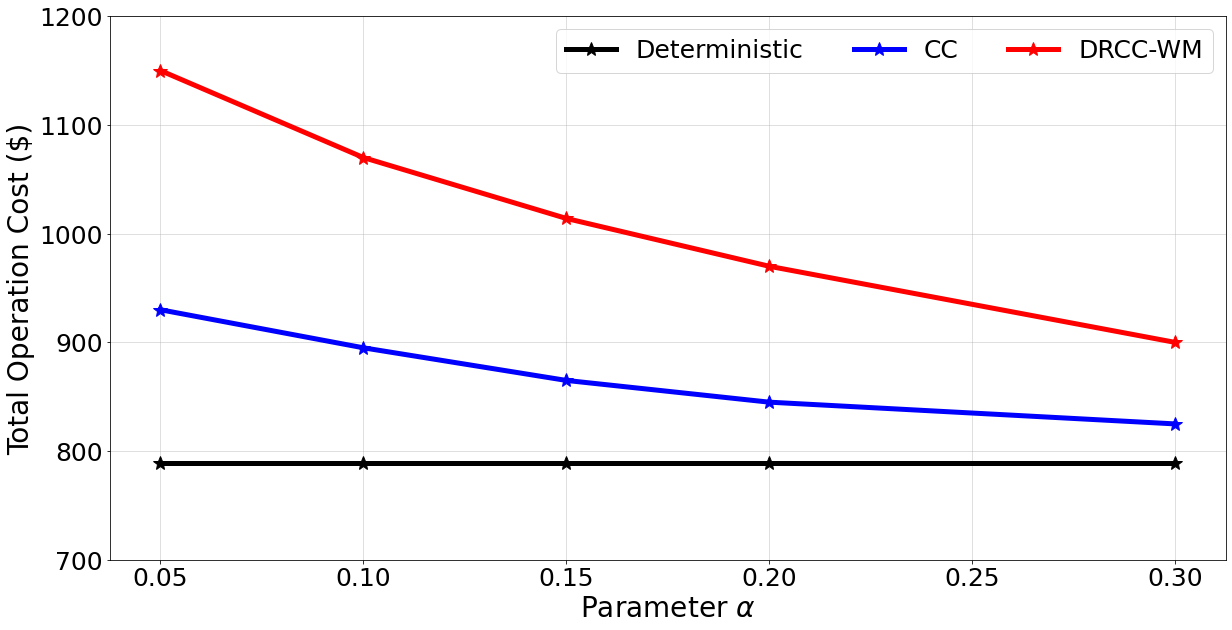}
	\caption{Operation cost sensitivity with different parameter $\alpha$ values (risk level).}
    \label{DRCCsensitivity}
\end{figure}

Figure \ref{DRCCsensitivity} shows the operational cost sensitivity for different parameter $\alpha$ values which is risk level. As it can be seen, the deterministic approach, represented by the black line, does not account for PV uncertainty and results in a constant, lower operational cost. This is because it assumes no variability or risk in PV output. The blue line corresponds to the Chance-Constrained (CC) method, which includes PV uncertainty by providing a fixed probabilistic guarantee. This leads to slightly higher costs compared to the deterministic approach, with costs decreasing as $\alpha$ increases. The red line represents the DRCC-WM, which accounts for worst-case PV uncertainty within a defined ambiguity set. This method yields the highest costs due to its conservative approach to risk management. As $\alpha$ increases, indicating lower risk aversion, the costs for all methods decrease. However, the DRCC-WM consistently results in higher costs due to its robust treatment of PV uncertainty. This shows the trade-off of minimizing operational costs and managing risk in energy management systems.

\subsection{Resilience Analysis of Apartment Complex Against Outages with Different Battery Sizes}
Table \ref{resilience_table} shows the numerical results of the resilience analysis for different battery capacities. At 100\% battery capacity (800 kWh), the unserved energy demand is 1.17 MWh for residential main electricity consumption, 0.3 MWh for common areas in residential buildings, and 0.03 MWh for commercial purposes. In total, the resilience index for this scenario is 80.1\%, indicating that the battery can fully supply the nighttime load. The reason that it is not 100\% is that during the initial hours of the day, there is no cheap solar electricity available, necessitating the purchase of power from the grid. It is noteworthy that with this battery size, the apartment complex becomes grid-independent and achieves zero carbon emissions, which is crucial for the future energy transition. As smaller battery capacities are utilized, the resilience index decreases, and the complex is no longer grid-independent.

\begin{table}[]
\label{ResilienceTable}
\centering
\footnotesize 
\captionsetup{labelsep=space,font={footnotesize,sc}}
\caption{ \\ Resilience Analysis of Apartment Complex}
\label{resilience_table}
\begin{tabular}{|c|ccc|c|}
\hline
\multirow{2}{*}{Case} &
  \multicolumn{3}{c|}{Energy Not Supplied (MWh)} &
  \multirow{2}{*}{\begin{tabular}[c]{@{}c@{}}Resilience\\ Index \\ (\%)\end{tabular}} \\ \cline{2-4}
 &
  \multicolumn{1}{c|}{Residential} &
  \multicolumn{1}{c|}{\begin{tabular}[c]{@{}c@{}}Residential\\ Common\end{tabular}} &
  Commercial &
   \\ \hline
100\% Battery & \multicolumn{1}{c|}{1.17} & \multicolumn{1}{c|}{0.3}  & 0.03 & 80.1 \\ \hline
75\% Battery  & \multicolumn{1}{c|}{1.4}  & \multicolumn{1}{c|}{0.51} & 0.08 & 73.5 \\ \hline
50\% Battery  & \multicolumn{1}{c|}{1.98} & \multicolumn{1}{c|}{0.51} & 0.08 & 65.6 \\ \hline
\end{tabular}
\end{table}

\section{Conclusion}\label{conclusion}\label{conclusionn}
This paper proposed a resilience-focused energy management model for grid-connected residential apartment complexes equipped with PV systems, battery storage, and EV charging stations. By addressing PV power uncertainty through a distributionally robust chance-constrained approach using the Wasserstein metric, the model ensured reliable and cost-effective operation. Real-world data, including PV generation, electric loads, and pricing structures for an existing apartment complex in Orlando, was used to validate the approach. The results showed that integrating PV and battery systems reduced operational costs, lowered emissions, and supported EV adoption while offering a profitable business model for property owners. These findings highlight a practical and sustainable framework for advancing clean energy use in residential complexes. Future work will focus on incorporating additional uncertain parameters such as electric load and EV charging behavior. Machine learning-based optimization methods will also be explored to enhance the model’s scalability and decision-making under uncertainty.

\setstretch{0.9}

\bibliographystyle{IEEEtran}
\bibliography{mybib}

\end{document}